\documentclass[aps,pra,twocolumn,a4paper,groupedaddress,floatfix,10pt,noeprint]{revtex4-1}

\usepackage{graphicx,amsmath,amssymb,amsfonts,dsfont}
\newcommand{\mf}[1]{\boldsymbol{#1}}
\newcommand{\ket}[1]{\ensuremath{|#1\rangle}}
\newcommand{\mc}[1]{\ensuremath{\mathcal{#1}}}
\newcommand{\bra}[1]{\ensuremath{\langle #1 |}}
\newcommand{\braket}[2]{\ensuremath{\langle #1 | #2 \rangle}}

\newcommand{\imag}{\mathrm{i}}
\usepackage{color}

\newcommand{\rhoU}{\rho_U}

\DeclareMathAlphabet\mathbfcal{OMS}{cmsy}{b}{n}

\usepackage{mathtools,nicefrac}
\usepackage{float}
\DeclarePairedDelimiter{\evdel}{\langle}{\rangle}
\newcommand{\ex}{\operatorname{}\evdel}

\begin{document}
\title{Probing microscopic models for system-bath interactions via parametric driving}
\author{Anastasia S. D. Dietrich${}^{1}$}
\author{Martin Kiffner${}^{2,1}$}
\author{Dieter Jaksch${}^{1,2}$}

\affiliation{Clarendon Laboratory, University of Oxford, Parks Road, Oxford OX1
3PU, United Kingdom${}^1$}
\affiliation{Centre for Quantum Technologies, National University of Singapore,
3 Science Drive 2, Singapore 117543${}^2$}
\date{\today}

\begin{abstract}
We show that strong parametric driving of a quantum harmonic oscillator coupled to a thermal bath allows one 
to distinguish between different microscopic models for the oscillator-bath coupling. 
We consider a bath with an Ohmic spectral density and a model where the system-bath interaction can be tuned continuously 
between position and momentum coupling via the coupling angle $\alpha$. 
We derive a  master equation for the reduced density operator of the oscillator in Born-Markov approximation 
and investigate its quasi-steady state as a function of the driving parameters, the temperature of the bath and the coupling angle $\alpha$. 
We find that the time-averaged variance of position and momentum exhibits a strong dependence on these parameters. 
In particular, we identify parameter regimes that maximise the $\alpha$-dependence   and 
provide an intuitive explanation of our results.
\end{abstract}

\maketitle
\section{Introduction}
Accurate models for the interaction between a quantum system and its environment have been vital for 
the success  of quantum optics and cold atom systems. For example, they have 
paved the way for revolutionising methods such as laser cooling~\cite{metcalf:lct}, and 
are essential  for developing applications in quantum technologies where decoherence effects need to be as small as possible.  
For some systems our understanding of the system-bath coupling is so accurate that it can be used 
to control and engineer these interactions. In this way, the role of  dissipation and decoherence 
can be transformed from a detrimental effect into a wanted feature as shown, e.g., in~\cite{kiffner:10,kiffner:12d,Diehl2010,Honing2012,Diehl2008,Muller2012,Verstraete2009a}.

This is in stark contrast to  condensed matter systems where the exact microscopic model underlying the system-bath coupling 
is often unknown~\cite{Capan2003}, and thus one has to resort to phenomenological models. 
Driving these systems with intense terahertz radiation opens up unprecedented possibilities to manipulate 
their quantum dynamics~\cite{Mankowsky2017,Forst2014a,Forst2014,Nova2017,Caviglia2012,Forst2015b,Forst2017,Rini2007,Liu2012,Denny2015x,Schlawin2017,Fausti2011,Hu2014,Mitrano2016}. 
Examples include the melting of charge density waves~\cite{Mankowsky2017,Forst2014a,Forst2014}, the generation of synthetic magnetic fields~\cite{Nova2017}, 
the control of heterointerfaces~\cite{Caviglia2012,Forst2015b,Forst2017}, 
the possibility to drive metal-insulator transitions~\cite{Rini2007,Liu2012}, the parametric cooling of bilayer cuprate superconductors \cite{Denny2015x}, the control of transport modes in cuprate superconductors~\cite{Schlawin2017}, or even the controlled creation of
transient superconductivity~\cite{Fausti2011,Hu2014,Mitrano2016}.
In order to optimize the coherent control of these systems even further, it would be highly desirable to improve our understanding of their 
system-bath interactions. 

\begin{figure}[!t]
  \centering
\includegraphics[width=0.4\textwidth]{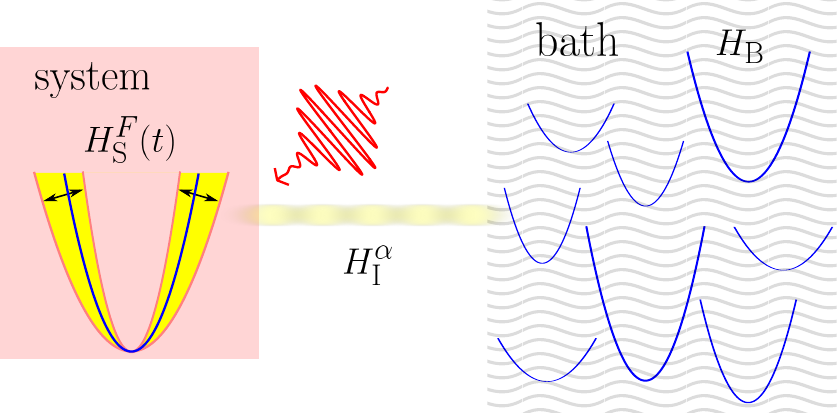}
    \caption{(Color online) The system of interest is comprised of a parametrically driven harmonic 
oscillator  and a bath of harmonic oscillators with Hamiltonians  $H_\text{S}^F(t)$ 
and $H_\text{B}$, respectively. The system-bath interaction is described by the interaction Hamiltonian $H^{\alpha}_I$, where the parameter $\alpha$ 
allows one to continuously adjust the interaction  from position ($\alpha=0$) 
to momentum coupling ($\alpha=\pi/2$). The parametric driving affects both the system and dissipation mechanism as indicated by the red arrow.}
    \label{Param_schematic}
\end{figure}

A direct manifestation of system-bath interactions are decay and decoherence rates that can be directly observed in an experiment. 
However, these quantities reveal little about the microscopic details of the system-bath interaction, and 
additional steps need to be taken to uncover more details. 
For example, some information about the local spectral density of a condensed matter heat bath can be obtained 
by observing  the non-Markovian behaviour of an opto-mechanical resonator coupled to it~\cite{Groblacher2015}. 
Alternatively,  information about the system-bath coupling can in principle be obtained by driving the system of interest so strongly that the driving 
modifies the system-bath interaction. This strongly-driven regime can be   described within the framework of the 
Floquet-Markov master equation approach~\cite{Graham1994,Breuer2002,Blumel1991,Kohler1997,Breuer1997,Donvil2017}, which 
combines ideas from open quantum systems with Floquet theory~\cite{Floquet1883,Shirley1965,Hanggi1997}. 
For example, it has been shown~\cite{Gasparinetti2013} that the response of a strongly driven two-level system depends on the specific form of the coupling operator. 
Here we consider a paradigmatic example for system-bath interactions comprised 
of a parametrically driven oscillator that is weakly coupled to a 
thermal bath, see Fig.~\ref{Param_schematic}. We show that the  driving 
allows one to distinguish between microscopic models for the system-bath interaction 
$H^{\alpha}_I$ that can be tuned  continuously 
from position to momentum coupling via the parameter $\alpha$. 
We derive a master equation for the reduced density operator of the oscillator in Born-Markov approximation and find 
that its quasi-steady state exhibits a strong dependence on $\alpha$ due to the parametric driving. 
In particular, we consider the time-averaged variance of position and momentum and 
identify parameter regimes that maximise their $\alpha$-dependence. 
We find that this $\alpha$-dependence  can be understood in terms of an effective model 
with an $\alpha$-dependent spectral density that is probed at different Floquet quasi-frequencies. 
This also shows that strong driving allows investigating other aspects of system-bath interactions 
that go beyond measuring the value of $\alpha$.

The paper is organised as follows. In Sec.~\ref{model} we present our model and discuss the steps and approximations leading to 
a master equation applicable to a strongly driven system. The presentation of our results in~Sec.~\ref{results} begins with a 
description of the observables that we use to characterise the quasi-steady state of the oscillator in Sec.~\ref{observables}. 
Our numerical results for the time-averaged  mean values of the system quadratures as a function of the coupling angle, bath temperature and driving 
parameters are presented in Sec.~\ref{numerics}. In Sec.~\ref{dependence}, 
we introduce a unitary transformation of the total Hamiltonian in order to explain the  $\alpha$-dependence of our results. 
Finally,  we conclude with a summary and an outlook for further work in Sec.~\ref{conclusion}.
\section{Model \label{model}}
Here we describe our theoretical model for a driven harmonic oscillator coupled 
to a thermal bath. The isolated system is described in Sec.~\ref{system}, and 
Sec.~\ref{sysbath} discusses the system-bath coupling. 
\subsection{Isolated system dynamics \label{system}}
We consider a harmonic oscillator with mass $m$ and time-dependent frequency 
$\omega_F(t)$. The corresponding Hamiltonian  is 
given by~\cite{Lo1991,Lo1993,Abdalla1993,Dantas1992a,Ji1995a,Sheng1995,Song1999}
\begin{equation}
\label{HSys}
H^F_{\text{S}}(t)=\frac{{p}^2}{2m}+\frac{1}{2}m\omega^2_F(t){x}^2,
\end{equation}
where  the position and momentum operators  $x$ and  $p$ obey the canonical 
commutation relation $[x,p]=\imag\hbar$. We assume that   the oscillator 
frequency 
$\omega_F(t)$ is modulated periodically with frequency $\omega_L$, 
\begin{align}
\omega^2_F(t)=\omega^2_0[1+F\cos{(\omega_L t+\phi_L)}],
\end{align}
where $F$ is the relative modulation amplitude. 
Throughout this paper we assume that $F=0$ for $t<0$. It follows that the 
Hamiltonian in Eq.~(\ref{HSys}) reduces to an ordinary harmonic oscillator
with frequency $\omega_0$ and mass $m$ for negative times. 
In our model the driving amplitude is suddenly switched on at $t=0$ and held at 
a constant value for $t>0$. 

The time-dependent  Schr\"odinger equation associated with the Hamiltonian in 
Eq.~(\ref{HSys}) can 
be solved exactly~\cite{Yeon1993,Song1999,Um2002,Ji1995,Pedrosa1997},
\begin{equation}
\imag\hbar\frac{\partial}{\partial_t}\psi_n(x,t)=H^F_{\text{S}}(t)\psi_n(x,t) 
\,.
\label{tdse}
\end{equation}
The solutions $\psi_n(x,t)$ ($n=0,1,2...$) are described in 
Appendix~\ref{exact} 
and reduce to the familiar harmonic oscillator states for $F=0$. It is possible 
to introduce generalized creation and annihilation operators 
that operate on the states  $\psi_n(x,t)$ like in the undriven harmonic 
oscillator~\cite{Kohler1997}, 
\begin{subequations}
 \label{RaisingLowering}
\begin{align}
A(t) \ket{\psi_{n}(t)}= & \sqrt{n}\ket{ \psi_{n-1}(t)},   \\
A^{\dagger}(t) \ket{ \psi_{n}(t)} = & \sqrt{n+1} \ket{\psi_{n +1}(t)}.  
\end{align}
\end{subequations}
Note that the operators $A$ and $A^{\dagger}$ are time-dependent. 
Time-dependent Hamiltonians  do generally not 
guarantee the validity of uncertainty relations 
and commutation relations~\cite{Um2002}. However, the solution to the isolated 
system dynamics for our system Hamiltonian in Eq.~(\ref{HSys})
does not  violate the uncertainty and 
commutation relations between canonical position and momentum 
operators~\cite{Pedrosa1997}.

\subsection{System-bath coupling \label{sysbath}}
We assume that the driven oscillator is weakly coupled to a heat bath of $N$ 
harmonic oscillators with Hamiltonian 
\begin{equation}
H_B=\sum_{r=1}^{N}\left(\frac{p^2_{r}}{2 m_{r}}+\frac{1}{2}m_{r} 
\omega^2_{r} x^2_{r}\right)\,.
\label{Hb}
\end{equation}
Here position and momentum operators of the $r$th  oscillator with mass $m_{r}$ 
and frequency $\omega_{r}$ 
are denoted by operators $x_{r}$ and $p_{r}$, respectively. 
The coupling between the bath and system is described by the  
Hamiltonian
\begin{equation}\label{EqInt}
H^{\alpha}_{\text{I}}=-c_{\alpha} B,
\end{equation}
where the parameter $\alpha$ in the coupling operator 
\begin{equation}
c_{\alpha}=\cos{(\alpha)}x+\sin{(\alpha)}\frac{p}{m \omega_0}
\label{coupling-op}
\end{equation}
describes which degrees of freedom are coupled to the bath. 
Note that $c_{\alpha=0}$ corresponds to position coupling, and 
$c_{\alpha=\pi/2}$ realises momentum coupling~\cite{Kohler2013,Bao2005,Leggett1984,Cuccoli2001}. 
The bath operator $B$ in Eq.~(\ref{EqInt}) is given by
\begin{align}
B=\sum_{r=1}^{N}\kappa_{r} x_{r} \,, 
 \label{Bdef}
\end{align}
and $\kappa_{r}$ are coupling constants. 
The system-bath interaction Hamiltonian in Eq.~(\ref{EqInt}) thus couples the system operator $c_{\alpha}$  to the position operators of the bath. 

With the definitions in Eqs.~(\ref{HSys}),~(\ref{Hb}) and~(\ref{EqInt}) we arrive at the  total Hamiltonian for the driven oscillator coupled to the bath, 
\begin{align}
 H^{\alpha}=H^F_S(t)+H_B+H^{\alpha}_I\,. 
 \label{Htot}
\end{align}
We assume the coupling to the bath to be weak and describe the quantum dynamics 
of the reduced density operator of the system $\rho$ by a Born-Markov master 
equation~\cite{Breuer2002}, 
\begin{align}
 \dot 
{\rho}=-\frac{\imag}{\hbar}\left[H^F_{\text{S}}(t),\rho\right]+\mathcal{K}\rho, 
\label{Master-Eq}
\end{align}
where the first term describes the coherent evolution and 
$\mathcal{K}\rho$ accounts for the  interaction with the heat bath at 
temperature $T$,
\begin{align}
\mathcal{K}{\rho}= & -\frac{1}{\hbar^2} 
\int\limits_0^{\infty}\text{d}\tau\text{Tr}_B\left\{\left[H^{\alpha}_{\text{I}},
\left[\tilde{H}^{\alpha}_{\text{I}}(t-\tau,t),\rho_B\otimes\rho(t)\right]\right] 
\right\}\,.
\label{BornMarkov}
\end{align}
Here 
\begin{align}
 \rho_B = \exp\left(-\frac{H_B}{k_{\text{B}} T}\right)/\text{Tr}_B 
\left\{\exp\left(-\frac{H_B}{k_{\text{B}} T}\right)\right\}
\end{align}
is the density operator of the bath in thermal equilibrium and   $k_{\text{B}}$ 
is Boltzmann's constant.
The operator $\tilde{H}^{\alpha}_{\text{I}}$ in Eq.~(\ref{BornMarkov}) is 
defined as 

\begin{align}
 \tilde{H}^{\alpha}_{\text{I}}(t,t_0) = 
\left[U_S(t,t_0)U_B(t,t_0)\right]^{\dagger} H^{\alpha}_{\text{I}} 
[U_S(t,t_0)U_B(t,t_0)],
\end{align}
where $U_{B}(t,t_0)=\exp[-(\imag/\hbar) H_B(t-t_0)]$ describes the free 
evolution of the bath. The time 
evolution operator of the system is 
\begin{equation}
U_{S}(t,t_0)=\left\{ 
\begin{array}{ll}
\mathcal{T}_{+}\exp{\left(-\frac{\imag}{\hbar} 
\int\limits^t_{t_0}\text{d}t'\left[H_{\text{S}}^F(t')\right]\right)}, & t\ge 
t_0, \\[0.5cm]
\mathcal{T}_-\exp{\left(\frac{\imag}{\hbar} 
\int\limits^{t_0}_{t}\text{d}t'\left[H_{\text{S}}^F(t')\right]\right)},&t<t_0 ,
\end{array}
\right.
\label{timesys}
\end{equation}
where $\mathcal{T}_+$ and $\mathcal{T}_-$  are the chronological and 
anti-chronological  time ordering operators, respectively.

In order to evaluate Eq.~(\ref{BornMarkov}) we later assume that the heat bath 
exhibits an Ohmic spectral density with a Lorentz-Drude cutoff 
$\Omega\gg \omega_0$~\cite{Breuer2002}, i.e.
\begin{equation}
J(\omega)=\frac{1}{\pi}\gamma \omega m \frac{\Omega^2}{\Omega^2+\omega^2},
\label{J_main}
\end{equation}
where $\gamma$ is the decay rate. 
With these assumptions 
Eq.~(\ref{BornMarkov}) can be written as (see Appendix~\ref{MasterDerivation})
\begin{align}
\mathcal{K}{\rho}= & \frac{1}{\hbar} \int\limits_{-\infty}^{\infty}  
\text{d}\omega J(\omega) n(\omega) \notag\\
& \times \int\limits_0^{\infty}\text{d}\tau\, e^{\imag\omega 
\tau}\left[\tilde{c}_{\alpha}(t-\tau,t){\rho},c_{\alpha}\right]\!+\!\text{H.c.},
\label{Eqdissip}
\end{align}
where $\text{H.c.}$ is the hermitian conjugate, 
\begin{align}
{n(\omega)=\frac{1}{e^{\nicefrac{\hbar \omega}{k_B T}}-1}}
\end{align}
is the Bose-Einstein distribution.
The operator $\tilde{c}_{\alpha}$ is defined as 
\begin{align}
\tilde{c}_{\alpha}(t,t_0)=\left[U_{\text{S}}(t,t_0)\right]^{\dagger}c_{\alpha}U_
{\text{S}}(t,t_0)
 \label{c-op}
\end{align}
 and depends on the driving amplitude $F$ via the time evolution operator in 
Eq.~(\ref{timesys}). 
 In order to simplify the master equation further we take advantage of the fact 
 that there exists a  complete set of exact solutions to 
the isolated system dynamics. The details of the derivation are given in 
Appendix~\ref{MasterDerivation}. We find 
\begin{align}
\dot {\rho}= & -\frac{\imag}{\hbar}\left[H^F_{\text{S}}(t),\rho\right]\notag \\
& +\Big\{ \mathcal{S}_1^{\alpha}(t) \left[A(t)\rho 
A^{\dagger}(t)-A^{\dagger}(t)A(t)\rho\right] \notag \\
& \qquad+\mathcal{S}_2^{\alpha}(t) \left[A^{\dagger}(t)\rho 
A(t)-A(t)A^{\dagger}(t)\rho\right] \notag \\
& \qquad+\mathcal{S}_3^{\alpha}(t) \left[A(t)\rho A(t)-A(t)A(t)\rho\right] 
\notag \\
& \qquad+\mathcal{S}_4^{\alpha}(t) \left[A^{\dagger}(t)\rho 
{A}^{\dagger}(t)-A^{\dagger}(t)A^{\dagger}(t)\rho\right] \notag \\
&  \qquad +\text{H.c.}\Big\}\,,
\label{eqMaster2}
\end{align}

where the time-dependent functions $\mathcal{S}_i^{\alpha}$ ($i\in\{1,2,3,4\})$ 
are defined in Appendix~\ref{MasterDerivation}. In order to be consistent with 
the separation of time-scales required by the Markov 
approximation~\cite{Weiss1999,Gardiner2000,Breuer2002}, the 
definitions of $\mathcal{S}_i^{\alpha}$ ($i\in\{1,2,3,4\})$ involve a time 
average to exclude any dynamic faster than the bath response time [see Eq. 
(\ref{SecRWA})]. In this way we avoid tracking all the degeneracies in the 
Floquet eigenenergies~\cite{Blumel1991,Hone2009}.

We show in Sec.~\ref{results} that  Eq.~(\ref{eqMaster2}) can be approximated by 
a simpler master equation in the high-temperature regime, 
which we refer to as high-temperature master equation (HTME). The HTME is obtained 
by  ignoring the driving term in Eq.~(\ref{c-op}) which results in 
\begin{align}
\tilde{c}_{\alpha}(t,t_0)\rightarrow 
e^{\frac{\imag}{\hbar}H_{\text{S}}^{0}(t-t_0)} c_{\alpha} 
e^{-\frac{\imag}{\hbar}H_{\text{S}}^{0}(t-t_0)}\,.
 \label{c-op2}
\end{align}

With this additional approximation, the HTME for the Ohmic bath can be written as 
\begin{align}
\dot {\rho}= & -\frac{\imag}{\hbar}\left[H^F_{\text{S}}(t),\rho\right]\notag \\
& + \left\{   \frac{\gamma}{2}n(\omega_0)\left[a^{\dagger} \rho, 
a\right]
+\frac{\gamma}{2}(n(\omega_0)+1)\left[a\rho,a^{\dagger}\right] 
\right. \notag \\
 &\qquad 
+\frac{\gamma}{2}n(\omega_0)e^{\imag2\alpha}\left[a^{\dagger}\rho, 
a^{\dagger}\right] \notag \\
 &\qquad  \left. 
+\frac{\gamma}{2}(n(\omega_0)+1)e^{-\imag2\alpha}\left[a\rho 
,a\right] + \text{H.c.}\right\}\,, 
\label{SimpleM}
\end{align}
where $a^{\dagger}$ and $a$ are the familiar creation and 
annihilation operators of the undriven 
harmonic oscillator. 
For an undriven system ($F=0$) Eq.~(\ref{SimpleM}) reduces to the standard 
quantum optical master equation 
after a rotating wave approximation that eliminates all the terms in the third and forth lines.\par 

\section{Results \label{results}}
\begin{figure}[!t]
  \centering
\includegraphics[width=0.45\textwidth]{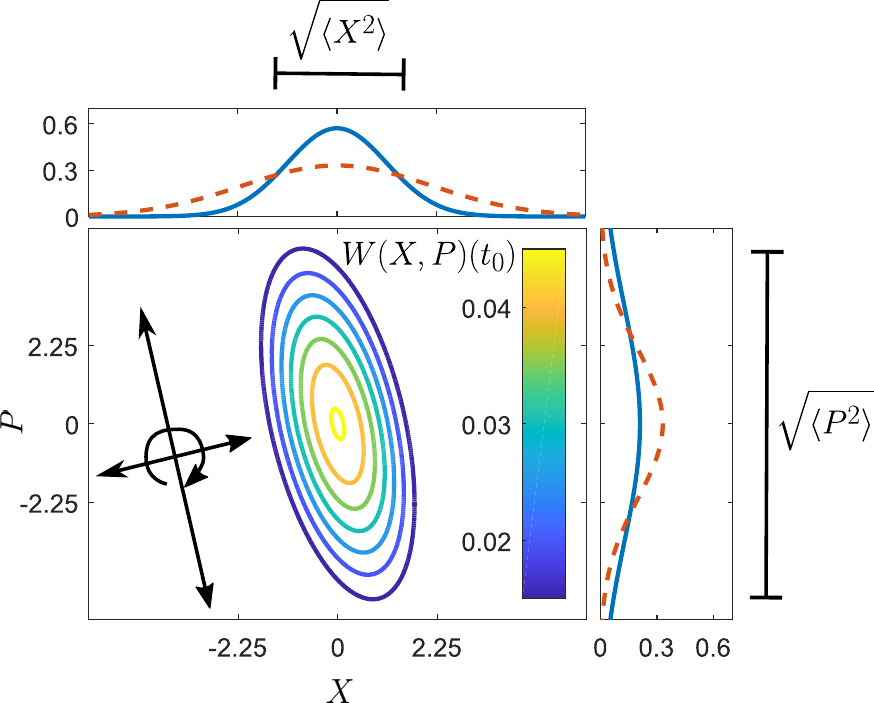}
    \caption{(Color online) The central panel shows a snapshot of $W(X,P)$ at time $t_0\gg 
{1}/{\gamma}$. The black arrow next to the ellipse indicates its dynamics 
consisting of simultaneous squeezing and rotation. The top and right panel 
contain the resulting probability distributions of the quadratures $X$ and $P$ 
(blue solid line). The red dashed line in the side panels corresponds to the 
initial thermal state for $t<0$. The width of the probability distribution is 
given by $\sqrt{\langle Y^2\rangle}$ with $Y=X,P$. Driving 
parameters are $b=0.1$, $q=0.5$ and $\zeta=0.7$ as defined in the text.} 
    \label{Wigner_schematic}
\end{figure}
Here we present a systematic study of the quasi-steady state of the driven harmonic 
oscillator as a function of the external driving parameters, the temperature of the 
bath and the coupling angle $\alpha$ in the system-bath Hamiltonian. 
In a first step, we describe how we characterise the quasi-steady state of the oscillator 
in Sec.~\ref{observables}. The numerical results are shown in Sec.~\ref{numerics}.

We introduce dimensionless parameters which are frequently used to 
characterise periodically driven systems~\cite{Bukov2015,Leibfried2003,Abramowitz1972}. For the specific model at hand, 
${b=\left(2\omega_0/\omega_L\right)^2}$ is proportional to the ratio of 
natural and driving frequencies squared and ${q=2F\omega^2_0/\omega^2_L}$  
characterises the effective driving strength. Throughout this work, we set the 
decay rate to ${\gamma/\omega_0=0.02}$ in order to be consistent with the 
Born Markov approximation. Note that $\gamma$ enters the master equation~(\ref{eqMaster2}) implicitly 
via the spectral density in Eq.~(\ref{J_main}). We further define ${\zeta=\hbar \omega_0/(k_B 
T)}$, which is proportional to the inverse temperature. 
\subsection{Observables \label{observables}}
 A descriptive way to illustrate the 
quantum state of the oscillator is given by the Wigner quasi probability 
distribution defined as \cite{Wigner1932,Polkovnikov2010}
\begin{equation}
 W(X,P)(t)=\frac{1}{\pi}\int_{-\infty}^\infty \langle X+Y| {\rho}(t) |X-Y \rangle e^{-2iP Y/\hbar}\,dY \,. 
\end{equation}
In this equation,  the density operator $\rho(t)$ is determined by Eq. (\ref{eqMaster2}) and $X$ and $P$ are 
dimensionless position and momentum operators, respectively, 
\begin{subequations}
\begin{align}
 X = &   \sqrt{\frac{m\omega_0}{\hbar}} x \,,\\
 P = & \sqrt{\frac{1}{m \omega_0 \hbar}} p \,.
\end{align}
\end{subequations}
For $t< 0$ the system is in a thermal state, which corresponds to a circular 
quasi-probability distribution with equally sized position and momentum 
quadratures ${{\langle X^2\rangle}}$ and ${{\langle P^2\rangle}}$, respectively. 
Once the driving is switched on at $t=0$, the shape of the Wigner function 
changes in time. A snapshot of $W(X,P)(t_0)$ at ${t_0\gg 1/\gamma}$ is 
shown in Fig.~\ref{Wigner_schematic}. In contrast to the thermal distribution, 
$W(X,P)(t_0)$ has an elliptic shape with a tilted major axis. The value of 
${{\langle X^2\rangle}}$ (${{\langle P^2\rangle}}$) has decreased (increased) 
with respect to the thermal distribution. Even though this is just an example of 
the Wigner function at a particular point in time, the whole dynamics of 
${W(X,P)(t\gg 1/{\gamma})}$ can be described by an ellipse rotating around 
the origin with period ${4\pi/\omega_L}$. During one revolution the 
angular velocity of 
rotation changes in time, and the ellipse grows and shrinks along its major and 
minor axis. 
\begin{figure}[!t]
  \centering
\includegraphics[width=0.45\textwidth]{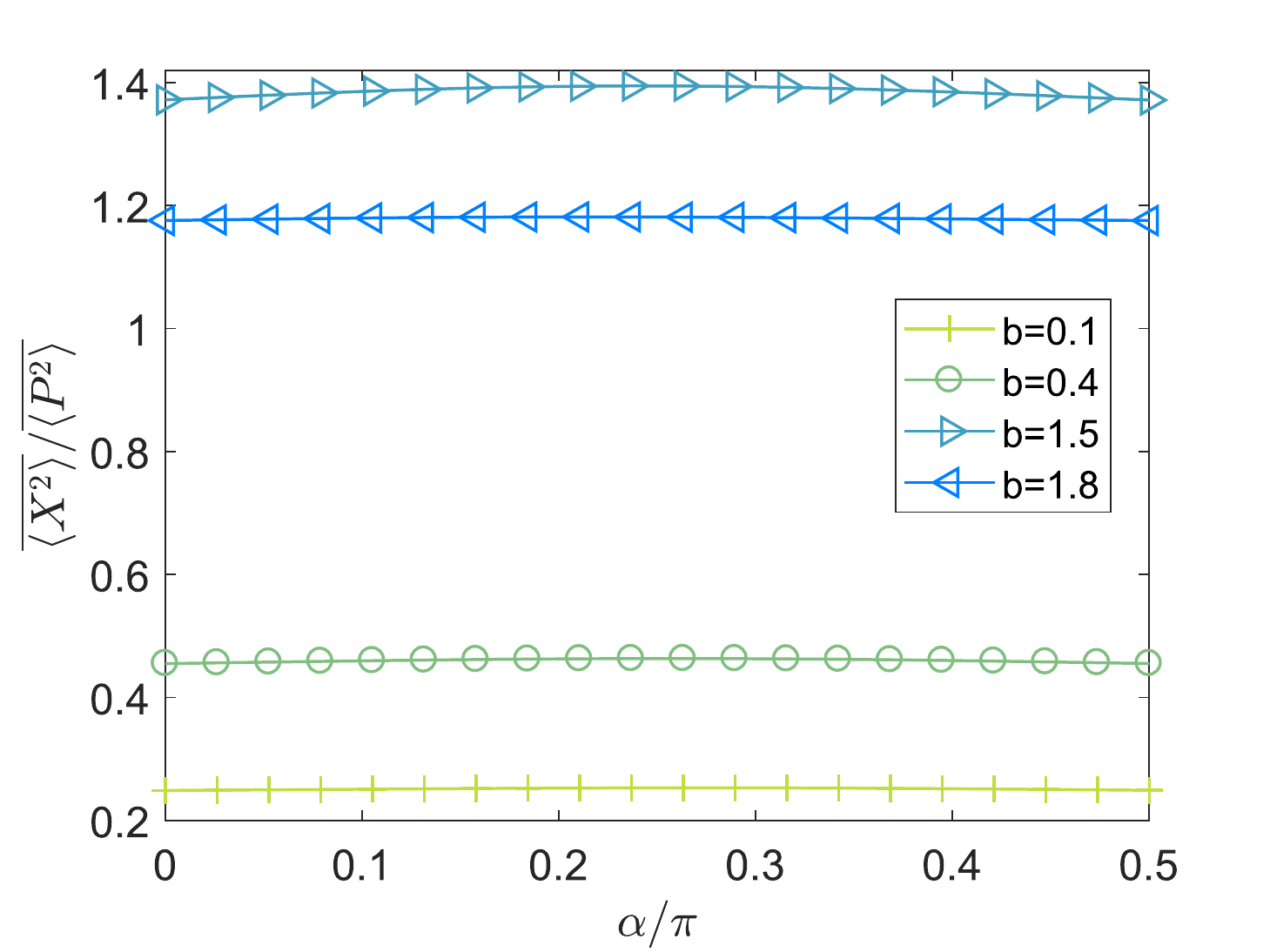}
    \caption{(Color online) Ratio of the time averaged quadratures $\overline{\langle X^2\rangle}/\overline{\langle P^2\rangle}$ 
    as a function of the coupling angle $\alpha$ and for different values of a. We have chosen $q=0.5$, $\zeta=10^{-4}$.}
    \label{alpha_idxrat}
\end{figure}

The dynamics of $\{\langle X^2\rangle ,\langle P^2\rangle,\langle X P+PX\rangle\}$ 
can be obtained from a closed set of equations since the total Hamiltonian  $H^{\alpha}$ in Eq.~(\ref{Htot}) 
is quadratic and therefore preserves the Gaussian nature of the initial thermal state~\cite{Zerbe1995,Rezek2006,Eisert2007,Galve2010,Chang2010,Schmidt2011}. This set 
of equations is derived from the full master equation~(\ref{eqMaster2}) and 
given in Appendix~\ref{FullEqu}. A similar closed set of equations can be obtained from 
the HTME in Eq.~(\ref{SimpleM}), and the corresponding results are denoted by $\langle\ \ldots \rangle_{\text{HTME}}$. 
In the following we consider the steady state 
regime ${t\gg 1/\gamma}$ and restrict our analysis to $\overline{\langle X^2\rangle}$ and $\overline{\langle P^2\rangle}$, where the bar indicates 
a time average over the interval ${\Delta t\approx 1/\gamma\gg 
2\pi/\omega_L}$. It follows that $\sqrt{\overline{\langle X^2\rangle}}$ 
and $\sqrt{\overline{\langle P^2\rangle}}$ characterise the width of the 
time-averaged Wigner function in $X$ and $P$, respectively.\par 
\subsection{Numerical results \label{numerics}}
\begin{figure}[!t]
  \centering
\includegraphics[width=0.49\textwidth]{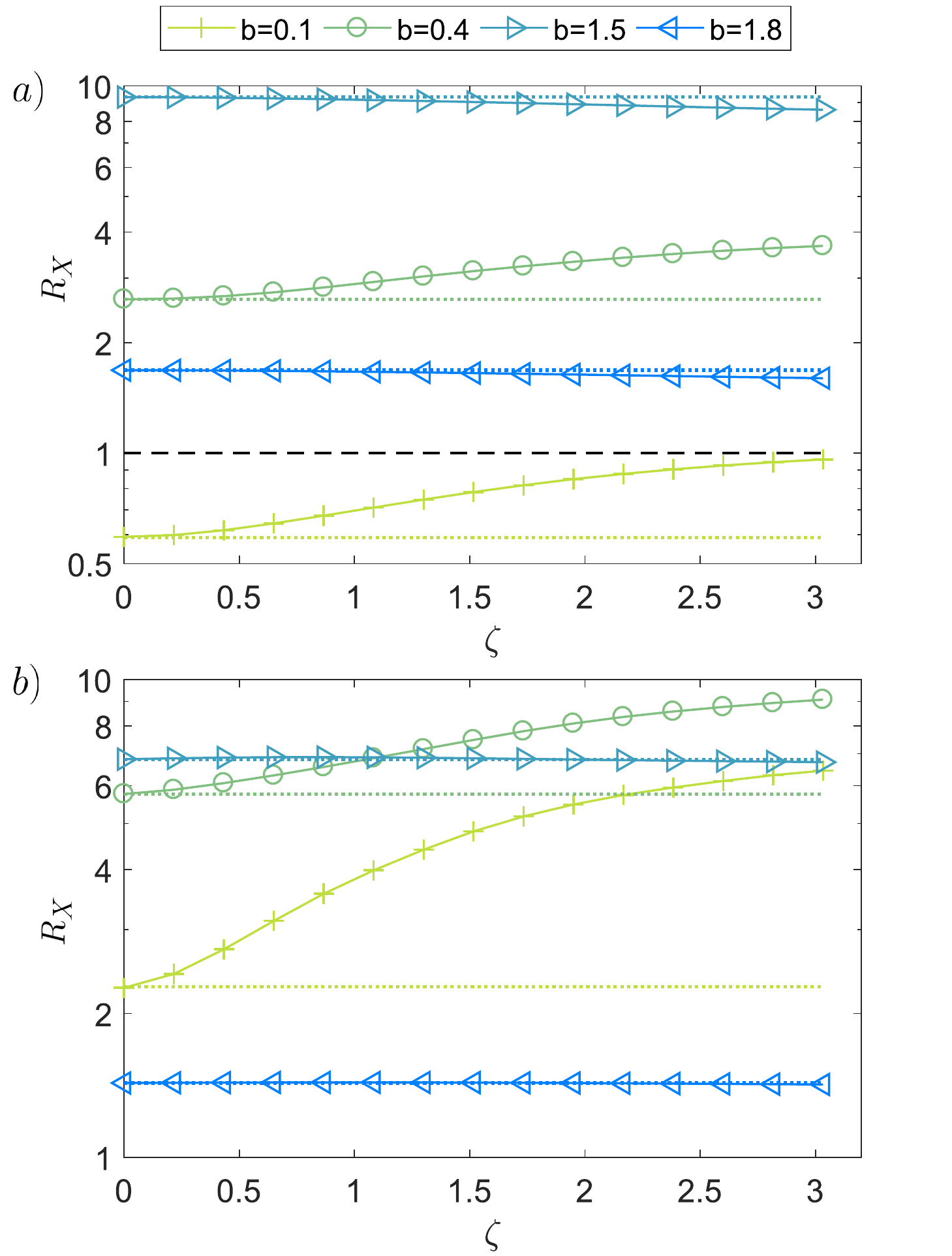}
    \caption{(Color online) $R_X$ as a function of inverse temperature 
$\zeta$ for $q=0.5$, different driving frequencies and coupling angle a) 
$\alpha=0$ and b) $\alpha=\frac{\pi}{2}$. Dotted lines of the same color 
indicate the results obtained from the evolution of the HTME. Dashed line at 
$R_X=1$ in a) is a guide to the eye.}
    \label{Aidx1}
\end{figure}
Here we systematically investigate the dependence of 
$\overline{\langle X^2\rangle}$ and $\overline{\langle P^2\rangle}$ on the parameters 
$b$, $\zeta$ and $\alpha$. 
In a first step, we find that the ratio $\overline{\langle X^2\rangle}/\overline{\langle P^2\rangle}$ 
is approximately independent of the coupling angle $\alpha$ as 
shown in Fig.~\ref{alpha_idxrat}. This ratio only depends on the driving 
frequency and strength, and is larger (smaller) than unity for $b>1$ ($b<1$). 
The shape of the time-averaged Wigner function thus deviates significantly 
from a circle in the presence of the parametric driving. 
The results in Fig.~\ref{alpha_idxrat} correspond to the high-temperature limit 
$\zeta=10^{-4}$. However, we find that the ratio $\overline{\langle 
X^2\rangle}/\overline{\langle P^2\rangle}$ is also approximately independent 
of $\zeta$ for the considered range of values ${\zeta\in \left[10^{-4},3\right]}$. 
The previously discussed dependence of $\overline{\langle X^2\rangle}$ and 
$\overline{\langle P^2\rangle}$ allows us to restrict the following analysis to 
one of the two quadratures. We choose $\overline{\langle X^2\rangle}$ 
and investigate its dependence on $b$, $\alpha$ and $\zeta$ by 
introducing the short-hand notation
\begin{equation}
R_X=\overline{\langle X^2\rangle}/\langle
X^2\rangle_{\text{thermal}},
\end{equation} 
where $\langle X^2\rangle_{\text{thermal}}$ is the undriven initial thermal value of 
$\overline{\langle X^2\rangle}$. A value of $R_X>1$ [$R_X<1$] 
thus means that  the driving enhances [reduces] the value of $\overline{\langle X^2\rangle}$ compared 
to its thermal value. 

\begin{figure}[!t]
  \centering
\includegraphics[width=0.49\textwidth]{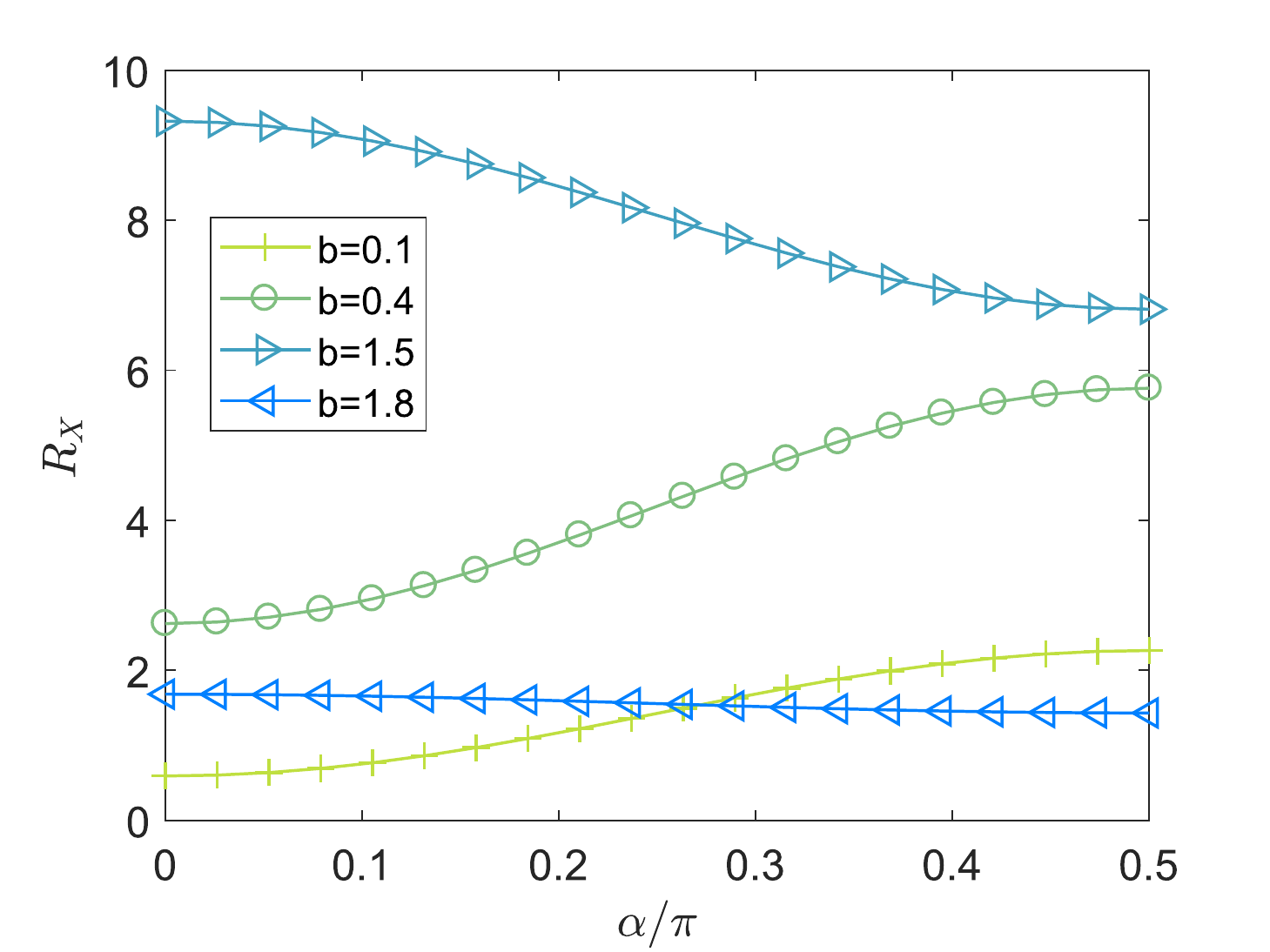}
    \caption{(Color online) ${R_X}$ as a function of coupling angle $\alpha$ 
for different values of a, driving strength $q=0.5$ and inverse temperature 
$\zeta=10^{-4}$.}
    \label{alpha_idx0}
\end{figure}
The dependence of   $R_X$ on  $\zeta$ is shown  in Figs.~\ref{Aidx1}a) and~b) for position   
and momentum coupling, respectively. Note that all values for $R_X$ in Fig.~\ref{Aidx1} differ from 
unity and hence the parametric driving  changes the 
value of $\overline{\langle X^2\rangle}$ compared to $\langle X^2\rangle_{\text{thermal}}$. 
Furthermore, we note that  the results for $R_X$ in Fig.~\ref{Aidx1}  
differ significantly for position and momentum coupling. Most prominently, $R_X>1$ for all curves in Fig.~\ref{Aidx1}b) (momentum coupling), 
while $R_X < 1$ for position coupling and  $b=0.1$. 
The reduction of $(R_X)_{\zeta \rightarrow 0}$ below unity for position coupling and fast driving ($b\ll1$) 
has already been studied in Refs.~\cite{Blatt1986,Zerbe1994a,Zerbe1995,Arnold1995}. 
In general, we find  that  $R_X$ strongly increases with $\zeta$ for both coupling types if $b<1$. 
On the other hand, $R_X$ shows a much weaker dependence on $\zeta$ 
for $b>1$. The value of $(R_X)_{b>1}$ slightly 
decreases with $\zeta$ for position coupling and remains approximately constant for momentum coupling. 
In order to illustrate the significance of the $\zeta$-dependence of $R_X$, we show 
the results for $(R_X)_{\text{HTME}}$ obtained via 
the HTME in Eq.~(\ref{SimpleM})   by the horizontal dotted lines in Fig.~\ref{Aidx1}. 
Since they exhibit no $\zeta$-dependence, the temperature dependence of $R_X$ predicted by 
the full master equation is a direct consequence of the parametric driving on the system-bath 
coupling. The differences between the HTME and full master equation are most pronounced near $\zeta=3$ and for $b=0.1$. 
On the other hand, we find that $R_X$ converges to $(R_X)_{\text{HTME}}$  in the limit $\zeta\rightarrow 0$ 
which can be understood as follows. The 
thermal correlation time of the bath $\tau_B=\hbar/(2\pi k_B T)$~\cite{Gardiner2000,Breuer2002} becomes much smaller than the typical timescale 
 $\tau_{\text{S}}=\text{min}(1/\omega_0,1/\omega_L)$ of  the 
 system dynamics in the high-temperature limit. Consequently, the time-evolution of the system 
 cannot influence the system-bath coupling.

Next we investigate the $\alpha$-dependence of $R_X$ in the high-temperature limit and for different driving 
frequencies. The results are  shown in Fig.~\ref{alpha_idx0}. We observe an 
increase (decrease) of $R_X$ with increasing $\alpha$ for 
$b<1$ ($b>1$). This behaviour is similar to the functional dependence of 
$R_X$ on $\zeta$ and $b$ shown in Fig.~\ref{Aidx1}. We thus find that increasing $\alpha$ or $\zeta$ leads to qualitatively comparable 
effects.
Most importantly,   $R_X$ shows a strong dependence on $\alpha$ for $b=0.4$ and $b=1.5$. Since 
$R_X$  is monotonously increasing (decreasing) for  $b=0.4$ ($b=1.5$), 
one can  determine the coupling angle $\alpha$ by comparing $\overline{\langle X^2\rangle}$ to $\langle X^2\rangle_{\text{thermal}}$. 
\begin{figure}[!t]
  \centering
\includegraphics[width=0.5\textwidth]{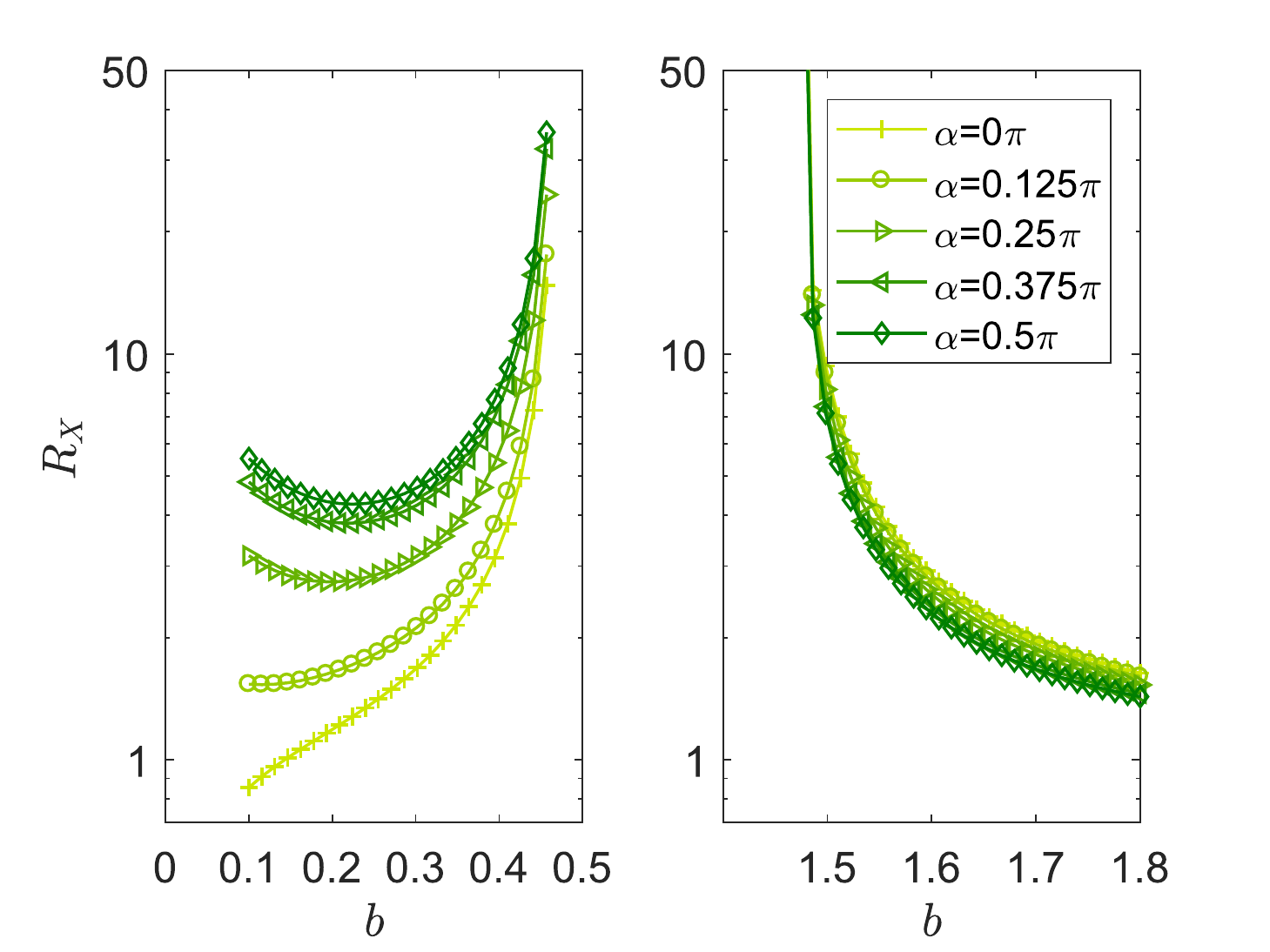}
\caption{(Color online) ${R_X}$ as a function of inverse driving frequencies 
for different coupling angles $\alpha$, driving strength $q=0.5$ and inverse 
temperature $\zeta=2$. The left and right panel correspond to driving above and 
below the resonance frequency, respectively.}
    \label{idx52blowT}
\end{figure}

The $\alpha$-dependence of the results in Fig.~\ref{alpha_idx0} are a direct consequence of the counter-rotating terms 
$\propto e^{\pm2\imag\alpha}$ in the HTME in Eq.~(\ref{SimpleM}).  
If the counter-rotating terms in the HTME are 
neglected within a rotating-wave approximation, it is independent of  $\alpha$. 
Such a rotating-wave approximation is justified, e.g., for the undriven oscillator and hence we 
conclude that the $\alpha$-dependence is a direct consequence of the driving.

In order to identify the  parameter regime with the strongest $\alpha$-dependence, we move away from the  high-temperature limit and 
show  $R_X$ as a function of $b$ for different values of $\alpha$ in Fig.~\ref{idx52blowT}. Here the left and right panels correspond to 
the parameter range $b<0.5$ and $b>1.45$, respectively. For values closer to 
$b=1$ the solutions become unstable and no steady state can be achieved. Figure~\ref{idx52blowT} shows that $R_X$ exhibits the strongest 
dependence on $\alpha$ for  $b\approx 0.1$. In particular, note that the range of the $R_X$ values from $\alpha=0$  
to $\alpha=\pi/2$ is larger than in the corresponding curve for $b=0.1$ in Fig.~\ref{alpha_idx0} 
due to the different $\zeta$ values. 

\begin{figure}[!t]
  \centering
\includegraphics[width=0.45\textwidth]{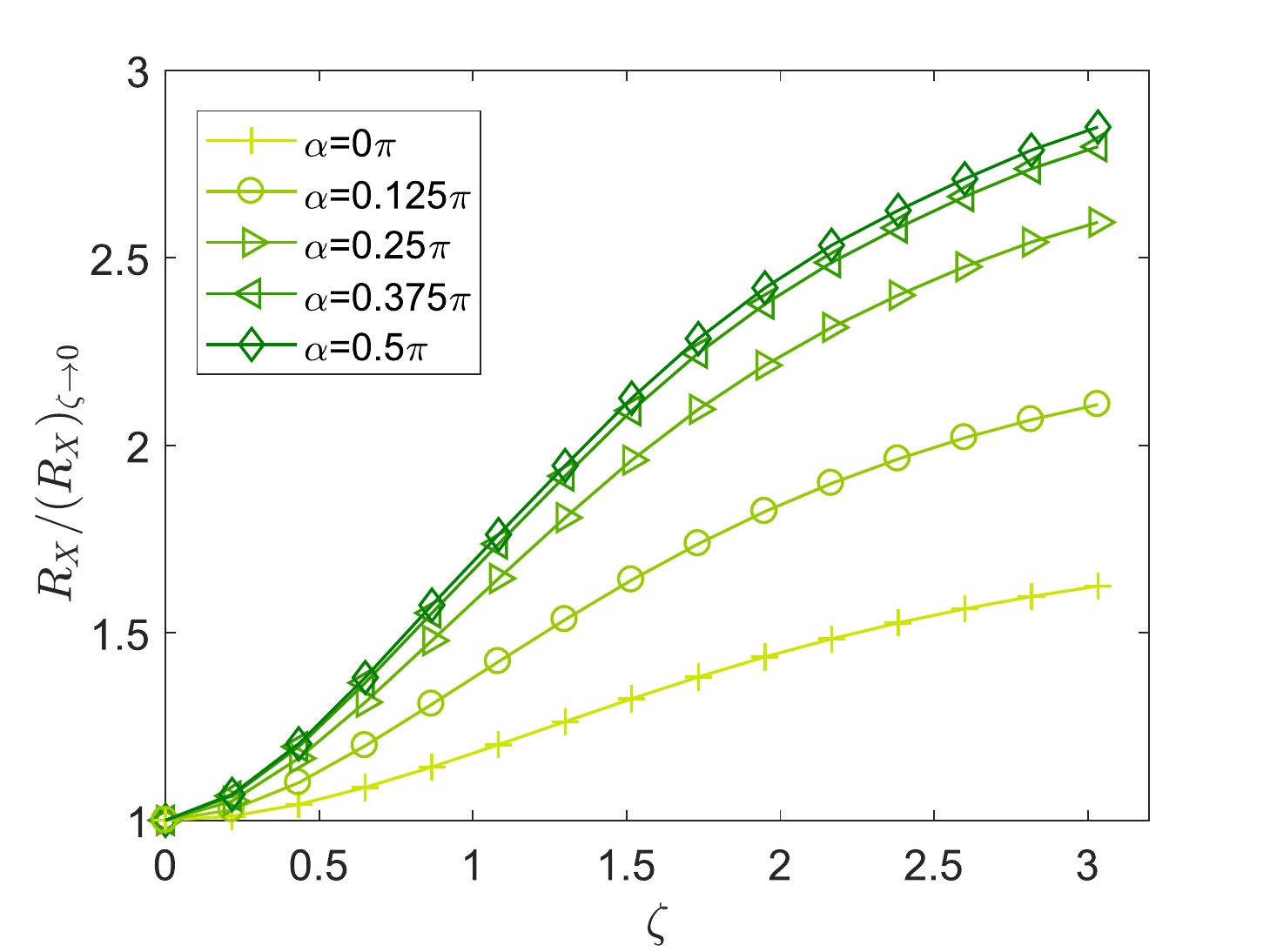}
\caption{(Color online) $R_X/(R_X)_{\zeta\rightarrow 0}$ as a function of $\zeta$ for $b=0.1$, $q=0.5$ 
and different values of $\alpha$. The values of 
$(R_X)_{\zeta\rightarrow 0}$ can be obtained from the $b=0.1$ curve in 
Fig. \ref{alpha_idx0}.}
    \label{zetaAreaa0.1}
\end{figure}

Finally, we focus on $b=0.1$ and investigate the $\zeta$-dependence of 
$R_X$  for different values of $\alpha$. 
We find that  $R_X$ increases relative to its  $\zeta \rightarrow 0$ 
value for all considered values of $\alpha$ as shown  in Fig.~\ref{zetaAreaa0.1}. 
The relative increase with $\zeta$ is the largest for 
$\alpha=\pi/2$ and the smallest for $\alpha=0$.
Furthermore, we find that small changes in $\alpha$ change the $\zeta$-dependence of 
$R_X/(R_X)_{\zeta\rightarrow 0}$ most significantly near $\alpha=0$.
Two conclusions can be drawn from the results in Fig.~\ref{zetaAreaa0.1}. First, 
increasing the value of $\zeta$ increases the spread of $R_X$ with $\alpha$. It thus becomes easier to 
determine $\alpha$ via a comparison of $\overline{\langle X^2\rangle}$ and $\langle X^2\rangle_{\text{thermal}}$ 
for smaller temperatures of the bath. Second, the 
difference of the $\zeta$-dependence 
of $R_X$ for various choices of $\alpha$ opens up a second route to determining $\alpha$. 
For example, one could increase $\zeta$ by a given factor and measure the corresponding increase in 
$R_X$. According to Fig.~\ref{zetaAreaa0.1}, this increase is a unique function of $\alpha$. 
Note that this approach works best  for small $\alpha$ where the $\zeta$-dependence 
of $R_X$ is most sensitive to changes in $\alpha$. 
\section{Discussion of $\alpha$-dependence \label{dependence}}
The aim of this section is to give an intuitive picture for the strong $\alpha$-dependence of 
the results in Sec.~\ref{numerics}. 
To this end we show in Appendix~\ref{SecInterH} that the 
total Hamiltonian $H^{\alpha}$ in Eq.~(\ref{Htot}) is unitarily equivalent to 
\begin{equation}
\mathcal{H}^{\alpha}=H^F_{\text{S}}(t)+H_{\text{B}}+\mathcal{H}^{\alpha}_{\text{I}} +H_{\text{shift}},
\label{Htrans}
\end{equation}
where $H^F_{\text{S}}(t)$ and $H_{\text{B}}$  remain unchanged and $H_{\text{shift}}$  is given in 
Appendix~\ref{SecInterH}. The latter term 
describes small shifts of the bath and system frequencies and will be  neglected  in the following.  
The transformed interaction Hamiltonian $\mathcal{H}^{\alpha}_{\text{I}}$ is given by
\begin{equation}\label{EqIntreplacalpha2}
\mathcal{H}^{\alpha}_{\text{I}}=-x B^{\alpha}\,,
\end{equation}
where 
\begin{align}
B^{\alpha}= \sum_{r=1}^{N}\kappa_r\left(\cos{(\alpha)}
x_r-\frac{\sin{(\alpha)}}{m_r \omega_0} p_{r}\right)\,. 
\label{balpha}
\end{align}
In contrast to $H^{\alpha}_{\text{I}}$ in  Eq.~(\ref{EqInt}), the system-bath coupling in 
$\mathcal{H}^{\alpha}_{\text{I}}$ 
is mediated by the position coordinate of the oscillator coupled to an $\alpha$-dependent 
superposition of position and momentum operators 
of the bath modes. In this way,  the $\alpha$-dependence has been entirely moved from the 
coupling operator to the bath. 

\begin{figure}[!t]
  \centering
\includegraphics[width=0.45\textwidth]{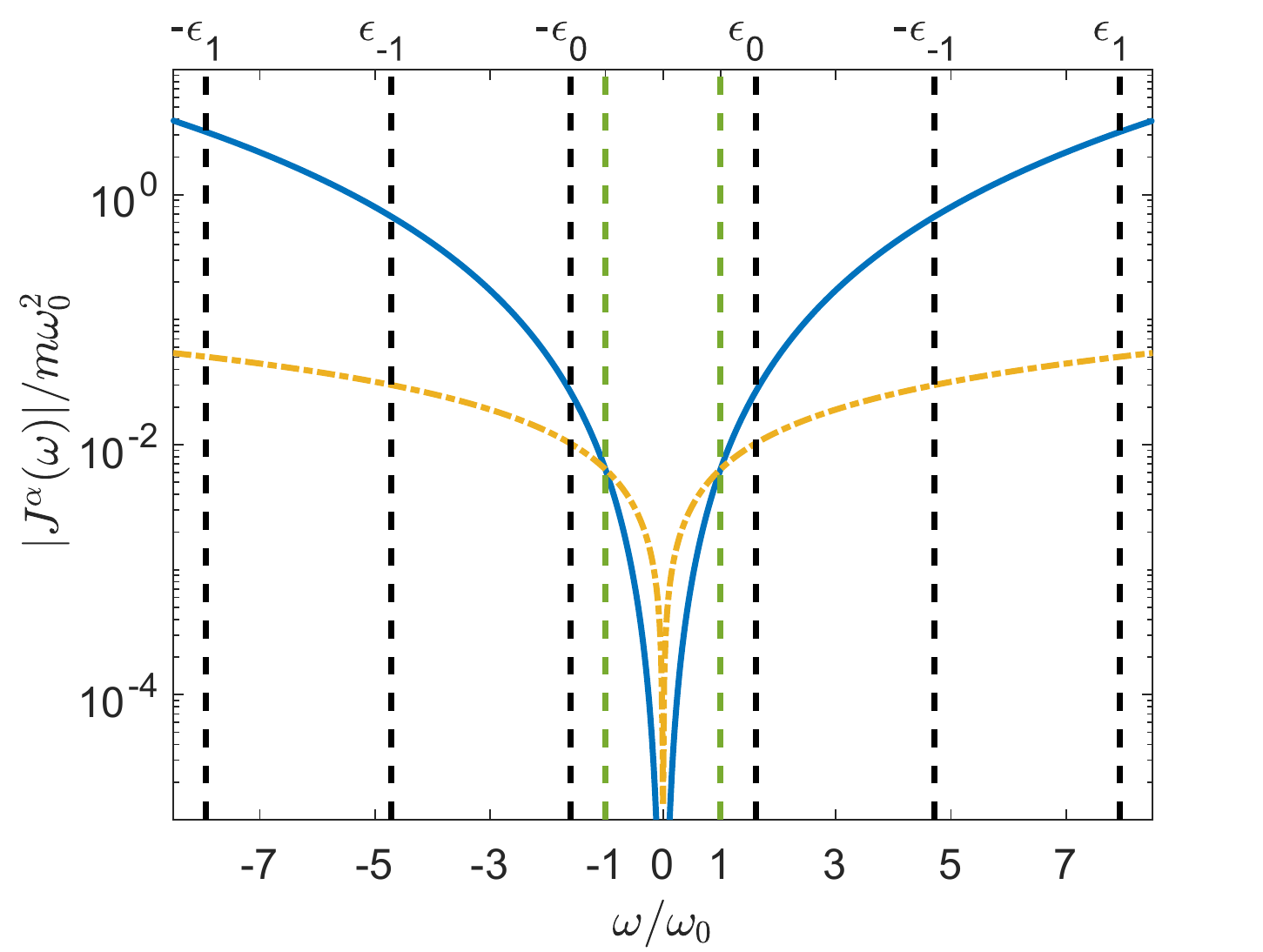}
    \caption{(Color online) The dash-dotted yellow (solid blue) line corresponds to the bath spectral 
density $J^{\alpha=0}(\omega)$ [$J^{\alpha=\pi/2}(\omega)$]. 
The green dashed vertical lines mark the undriven system 
frequency $\pm\omega_0$. The scaled Floquet quasi-frequencies $\epsilon_r=(\mu+r\omega_L)/\omega_0$ for $q=0.5$ and  
$b=0.1$ are indicated by black dashed  vertical lines.  }
    \label{spectr1}
\end{figure}

Following the steps in Appendix~\ref{MasterDerivation} we derive a master equation 
for the density operator $\rhoU$ from the transformed 
Hamiltonian $\mathcal{H}^{\alpha}$, see Appendix~\ref{SecInterH}.
We have explicitly verified that  $\rhoU$ and $\rho$ lead to the same results for all parameters investigated in Sec.~\ref{numerics}. 
In the derivation of the master equation for $\rhoU$, the linear superposition of bath operators in Eq.~(\ref{balpha}) gives rise to an $\alpha$-dependent spectral density
\begin{equation}
J^{\alpha}(\omega)=J(\omega)\left[\cos{(\alpha)}^2+\left(\frac{\omega}{\omega_0}\right)^2\sin^2{(\alpha)}\right],
\label{J2}
\end{equation}
which is the product of the  Ohmic spectral density $J(\omega)$  in~Eq.~(\ref{J_main}) 
and an  $\alpha$-dependent function. 
We show  $J^{\alpha}(\omega)$  for the two extreme cases $\alpha=0$ and $\alpha=\pi/2$ 
in Fig.~\ref{spectr1}. While $J^{0}(\omega)$ and $J^{\pi/2}(\omega)$ are identical at $\omega=\pm \omega_0$,  they 
differ significantly for $|\omega|\neq\omega_0$.
We are now in a position to understand the $\alpha$-dependence of the results in Sec.~\ref{numerics}. 
In general, the system-bath interaction is determined by the values of $J^{\alpha}(\omega_c)$ at  the 
characteristic frequencies $\omega_c$ of the  classical solution  of the  undamped harmonic 
oscillator [see Appendix~\ref{MasterU} for details].  
We have found that all observables in Sec.~\ref{numerics} are independent of $\alpha$ for an undriven oscillator. In this case, 
the resonance frequency $\omega_0$ is the only characteristic  frequency involved in the system dynamics. 
However, the  spectral density in Eq.~(\ref{J2}) satisfies $J^{\alpha}(\pm \omega_0)=J(\pm\omega_0)$ for all $\alpha$, and hence 
the system-bath interaction is independent of $\alpha$. 

This picture changes significantly if the driving is switched on. 
The driving leads to new characteristic frequencies  $\omega_c=\pm \mu+r\omega_L$, 
with $r\in \mathbb{Z}$ and where $\mu$ is the quasi-frequency of the Floquet spectrum closest to $\omega_0$ 
of the underlying classical model of a driven but undamped harmonic oscillator. 
Since the spectral density $J^{\alpha}$ 
depends strongly on $\alpha$ for $|\omega| > \omega_0$ [see Fig.~\ref{spectr1}],  
the parametric driving leads  to $\alpha$-dependent results.
\section{Summary and Conclusion \label{conclusion}}
We have shown that parametric driving of a harmonic oscillator coupled to a heat bath allows one 
to distinguish between different microscopic models for the system-bath interaction. 
More specifically, we have considered a bath with an Ohmic spectral density that couples 
to a linear superposition of the position and momentum degrees of freedom of the oscillator. 
This superposition is parametrised via the coupling angle $\alpha$ which allows one to continuously 
change the character of the coupling from position to momentum coupling. 
We have systematically investigated the dependence of the time-averaged expectation values of $\overline{\langle X^2\rangle}$ 
and $\overline{\langle P^2\rangle}$ on the driving parameters, temperature $T$ of the bath and $\alpha$. 
While $\overline{\langle X^2\rangle}/\overline{\langle P^2\rangle}$ is 
approximately independent of $\alpha$ and $T$, we have shown that 
$R_X=\overline{\langle X^2\rangle}/\langle X^2\rangle_{\text{thermal}}$ 
shows a strong dependence on temperature and on the coupling angle $\alpha$. 
This dependence on the coupling angle could be used to determine $\alpha$ by measuring $\overline{\langle X^2\rangle}$  
($\langle X^2\rangle_{\text{thermal}}$ in the presence (absence) of the parametric driving. 
We have found that $R_X$ exhibits the strongest dependence on  $\alpha$ for large driving frequencies $\omega_L\gg\omega_0$ and 
for bath temperatures with $k_B T \le \hbar \omega_0$. 
In  this regime, $R_X$ also displays a characteristic dependence on inverse temperature for each value of $\alpha$, thus 
offering an additional route to determine the value of $\alpha$ in the interaction Hamiltonian.

Our results have been obtained within the framework of a master equation in Born-Markov approximation that accounts for 
the modification of the system-bath interaction due to the parametric driving. 
In order to be consistent with the Markov approximation, we describe the time evolution of the reduced density operator 
of the oscillator on a coarse-grained time axis with resolution $\Delta t$, where $\Delta t$ is much smaller than $1/\gamma$ and  
much larger than the bath correlation time. 
We compare our results to a simpler master equation that neglects the influence of the driving on the dissipative part and 
where all counter-rotating terms are kept. We find that this HTME agrees with the full master equation 
in the high-temperature limit where $\hbar \omega_0 \ll k_B T$, but deviates significantly otherwise. 
We thus conclude that the influence of the parametric driving on the system-bath coupling must be taken into 
account in general.

We have shown that the  $\alpha$-dependence of our results can be understood  in an intuitive way by applying a unitary 
transformation to the total system-bath Hamiltonian. In this transformed picture, the 
position  variable of the oscillator couples to an $\alpha$-dependent superposition of bath variables which results in 
an $\alpha$-dependent spectral density $J^{\alpha}$. 
Since the system-bath coupling is determined by the values of $J^{\alpha}$ at the Floquet quasi-frequencies associated with the parametric driving, 
the density operator obtained from this master equation  depends on $\alpha$ if the system is driven. 

Our results open up several directions for future investigations. 
The explanation of the $\alpha$-dependence of our results via the unitarily equivalent model giving rise to an $\alpha$-dependent spectral density shows 
that strong parametric driving allows one to probe an unknown spectral density if the microscopic coupling mechanism between the system and the bath is known.
Furthermore, this approach could be extended to investigate systems with several competing dissipation mechanisms
or to the non-Markovian regime of strong damping which is often encountered in solid-state materials. 
This could either be achieved by using the Feynman-Vernon path integral formalism~\cite{Zerbe1995} or the stochastic Liouville-von Neumann equation~\cite{Stockburger2002}.
Eventually, an improved understanding of system-bath interactions in these systems paves the way 
to control and  engineer their quantum dynamics via strong driving. For example, cooling of a driven 
harmonic oscillator coupled to a non-Markovian bath can be achieved via optimal control techniques~\cite{Schmidt2011,Schmidt2012}. 
This could allow one to  prepare the oscillator in non-classical, squeezed states~\cite{Rahimi-Keshari2013,Sabapathy2016,Chen2018,Arvind1997}.

\begin{acknowledgments}
MK thanks the National Research Foundation and the Ministry of Education of 
Singapore for support. 
The research leading to these results has received
funding from the European Research Council under the European Unionʼs Seventh
Framework Programme (FP7/2007-2013)/ERC Grant Agreement no. 319286 Q-MAC 
and from EPSRC programme grant EP/P009565/1 DesOEQ.
\end{acknowledgments}
\appendix
\section{Exact solution of the driven harmonic oscillator \label{exact}}
Here we describe the exact solutions $\psi_n(x,t)$  to the 
time-dependent  Schr\"odinger equation in Eq.~(\ref{tdse}) following the 
apporach in~\cite{Ji1995,Pedrosa1997} that is 
based on dynamical invariants~\cite{Lewis1969,Ji1995a}. We write the solutions 
as 
\begin{align}
\psi_n(x,t)= & \frac{1}{\sqrt{2^n n!}}\left(\frac{{m}\omega_I}{\pi {\hbar} 
g_{-}(t)}\right)^{\frac{1}{4}}
e^{-\imag \frac{g_0(t)}{2 {\hbar}g_{-}(t)}x^2}e^{-i\Lambda(t)(n+\frac{1}{2})} 
\notag \\
& e^{-\frac{ {m}\omega_I}{2 {\hbar}g_{-}(t)}x^2}H_n\left(\sqrt{\frac{ 
{m}\omega_I}{{\hbar}g_{-}(t)}}x\right),
\label{wave}
\end{align}
where 
\begin{align}
\Lambda(t)=\int\limits_{0}^{t} \,\text{d}t'\frac{\omega_I}{g_{-}(t')},
\end{align}
is a global phase, 
\begin{align}
\omega_I={\frac{1}{m}} \sqrt{g_{+}(t)g_{-}(t)-g_0^2(t)}, 
\end{align}
is a  time-independent constant frequency and $H_n$ are the Hermite polynomials 
of degree $n$. 
The functions $g_{+}(t)$, $g_{+}(t)$ and $g_{+}(t)$ in Eq.~(\ref{wave}) are 
defined as 
\begin{subequations}
\begin{align}
g_{+}(t) & =m^2 \dot{f}_1(t)\dot{f}_2(t) , \\
g_0(t) & = -\frac{m}{2}[\dot{f}_1(t) f_2(t)+f_1(t)\dot{f}_2(t)] , \\
g_{-}(t) & =f_1(t)f_2(t), 
\end{align}
\end{subequations}
where  $f_{i}(t)$ ($i\in\{1,2\})$ are two linearly independent solutions  
of the classical harmonic oscillator. These functions obey
\begin{equation}
\ddot{f}_{i}(t) + \omega^2(t)f_{i}(t)=0,
\end{equation}
and we impose the initial condition $f_{i}(0)=1$ at $t=0$ and require 
$f_{1}(t)=f_2(t)^*$. 
For  $t<0$ where the driving is absent we thus find $f_{1,2}(t<0)=e^{\pm 
i\omega_0 t}$, 
and consequently  $g_{0}(t<0)=0$. Furthermore, as $g_{-}(t<0)=1$ the chosen 
initial conditions result in $\omega_I=\omega_0$ such that $\psi_n(x,t)$ reduces 
to the harmonic oscillator eigenstates. 
These states form an orthogonal basis and the states $\psi_n(x,t)$ remain 
orthogonal for $t\ge 0$, i.e. $\braket{\psi_n(t)}{\psi_m(t)}=\delta_{n,m}$ as 
the time dependence of the argument $\frac{m\omega_I}{\hbar g_{-}(t)}$ of the 
integral in the equal time norm can be eliminated by a change of variables. 
It follows that the states $\ket{\psi_n(t)}$ in Eq.~(\ref{wave}) span a complete 
basis at all times.
We  define generalised creation and annihilation operators $A^{\dagger}$ and $A$ 
that act on 
the states $\ket{\psi_n(t)}$ as described in Eq.~(\ref{RaisingLowering}). The 
annihilation 
operator is given by~\cite{Kohler1997}
\begin{align}
A(t) & 
=\frac{1}{\sqrt{2}}\left[{\sqrt{\frac{m\omega_0}{\hbar}}}h_1(t)x+{\sqrt{\frac{1}
{\hbar m\omega_0}}}h_2(t)p\right],
\label{A}
\end{align}
and the complex functions $h_1(t)$ and $h_2(t)$ are defined as
\begin{subequations}
\label{hfunc}
 \begin{align}
h_1(t) & 
=\exp{\left(\imag\Lambda(t)\right)}\sqrt{\frac{\omega_I}{{\omega_0}g_{-}(t)}}
\left(1+\imag\frac{ g_{0}(t)}{{m}\omega_I}\right),\\
h_2(t) & =\imag\,\exp{\left(\imag\Lambda(t)\right)}\sqrt{\frac{{\omega_0} 
g_{-}(t)}{\omega_I}}.
 \end{align}
\end{subequations}
Since $\left[x,p\right]=\imag\hbar$ and $h_1(t)h^*_2(t)-h^*_1(t)h_2(t)=-2\imag$, 
the 
equal-time commutation relation $\left[A(t),A^{\dagger}(t)\right]=1$ is 
satisfied. 
\section{Master equation derivation \label{MasterDerivation}}
Here we discuss the derivation of the master equation in Sec.~\ref{sysbath}. 
Our approach builds on the work presented in~\cite{Kohler1997}, but considers a  
more general system-bath coupling, does not use the Floquet basis states and employs the 
weakest possible rotating-wave approximation in order to be consistent with the Born-Markov 
approximation. 
In a first step we re-write the dissipative term  
$\mathcal{K}{\rho}$ in Eq.~(\ref{BornMarkov}) as 
follows~\cite{Breuer2002,Blumel1991,Hone2009}, 
\begin{align}
\mathcal{K}{\rho}= & \frac{1}{\hbar^2} 
 \int\limits_0^{\infty}\text{d}\tau \left(\frac{\imag}{2}D(\tau) 
[c_{\alpha},\{\tilde{c}_{\alpha}(t-\tau,t),\rho\}]\right.\notag \\
 &\qquad\qquad \left. -\frac{1}{2} 
D_1(\tau)[c_{\alpha},[\tilde{c}_{\alpha}(t-\tau,t),\rho]]\right),
\label{Intermediate}
\end{align}
where $\{Q,W\}=QW+WQ$ is the anti-commutator for operators $Q$, $W$ and we introduced following~\cite{Breuer2002} the dissipation kernel
\begin{equation}\label{kernel1}
D(\tau)=i\ex{[B,\tilde{B}(-\tau)]}=i\left[B,\tilde{B}(-\tau)\right],
\end{equation}
and the noise kernel
\begin{equation}\label{kernel2}
D_1(\tau)=\ex{\{B,\tilde{B}(-\tau)\}}\,. 
\end{equation}
We find 
\begin{subequations}
 \begin{align}
 \label{Dintegral}
 D(\tau) & = {2\hbar}\int\limits_0^{\infty}\text{d}\omega J(\omega)\sin 
\omega\tau \,,\\
 D_1(\tau)& = {2\hbar}\int\limits_0^{\infty}\text{d}\omega 
J(\omega)\coth(\hbar\omega/2k_B T)
 \cos\omega\tau\,,
 \label{Dintegral1}
 \end{align}
\end{subequations}
and 
\begin{align}
 J(\omega)=\sum_{n=1}^N\frac{\kappa^2_n}{2m_n \omega_n}\delta(\omega-\omega_n)
\end{align}
is the spectral density. As described in the 
main text we assume that $J(\omega)$ is Ohmic with a Lorentz-Drude cutoff as given in Eq.~(\ref{J_main}). 
The longest thermal correlation time is then given by 
$\tau_B=\text{max}[\hbar/(2\pi k_B T),\Omega^{-1}]$. We set $\Omega=10^5\omega_0$ such that 
for all considered parameters $\tau_B=\hbar/(2\pi k_B T)$. 
With the identities $J(\omega)=-J(-\omega)$ as well as $-n(-\omega)=n(\omega)+1$  we obtain Eq.~(\ref{Eqdissip}).  
The solution of the master equation~(\ref{Master-Eq}) with $\mathcal{K}{\rho}$ 
as in 
Eq.~(\ref{Eqdissip}) is greatly simplified by projecting it onto the states 
$\ket{\psi_{n}(t)}$ 
which solve the full Schr\"odinger equation of the driven oscillator. The matrix 
elements 
of the density operator of the system in this basis are denoted by 
\begin{align}
 \rho_{nm}(t) = & \bra{\psi_{n}(t)}\rho(t)\ket{\psi_{m}(t)}\,,
\end{align}
and evolve only due to the system-bath coupling. 
We find 
\begin{align}
\dot{\rho}_{nm}(t)= &
\sum_{k,l}\left[\bar{C}^{\alpha}_{nk}(t)\rho_{kl}(t)C^{\alpha}_{lm}(t)\right. 
\notag\\
& \left. \qquad 
-C^{\alpha}_{nk}(t)\bar{C}^{\alpha}_{kl}(t)\rho_{lm}(t)\right]\!+\!\text{H.c.},
\label{MasterEqProj}
\end{align}
where 
\begin{subequations}
\begin{align}
 C^{\alpha}_{nm}(t) = &\bra{\psi_{n}(t)}c_{\alpha}\ket{\psi_{m}(t)}\,, 
\label{c-mat}\\[0.3cm]
 \bar{C}^{\alpha}_{nm}(t)  =  & \frac{1}{\hbar}\int_{-\infty}^{\infty} 
\text{d}\omega \,J(\omega) n(\omega) \notag \\
  &\times \int_0^{\infty}\text{d}\tau\, e^{i\omega \tau}C^{\alpha}_{nm}(t-\tau) 
\,. \label{cbar-mat}
\end{align}
\end{subequations}
In order to evaluate these  matrix elements,  we 
express the coupling operator $c_{\alpha}$  
in Eq.~(\ref{coupling-op})  in terms 
of the generalised creation and annihilation operators  $A(t)$ and 
$A^{\dagger}(t)$, 
\begin{equation}
\label{calphaAAdagger}
c_{\alpha}=s^{\alpha}_1(t)A^{\dagger}(t)+s^{\alpha}_2(t)A(t),
\end{equation}
where 
\begin{equation}\label{Eqs1alpha}
s^{\alpha}_1(t)=\imag {\sqrt{\frac{\hbar}{2m \omega_0}}}\left[\sin{(\alpha)} 
h_1(t)-\cos{(\alpha)}h_2(t)\right]
\end{equation}
is a complex-valued function and $s^{\alpha}_2(t)=[s^{\alpha}_1(t)]^*$.
The matrix element in Eq.~(\ref{c-mat}) thus reads
\begin{equation}
C^{\alpha}_{nm}(t) = s^{\alpha}_1(t)A_{nm}^{\dagger} +s^{\alpha}_2(t)A_{nm},
\label{Cnm}
\end{equation}
and the time-independent matrix elements of the creation and annihilation 
operators are given by
\begin{subequations}
\begin{align}
  A_{nm}^{\dagger} &  = \bra{\psi_n(t)}A^{\dagger}(t)\ket{\psi_m(t)} = 
\sqrt{m+1}\delta_{n,m+1} \,,\\
 A_{nm}& = \bra{\psi_n(t)}A(t)\ket{\psi_m(t)} = \sqrt{m}\delta_{n,m-1}\,.
\end{align}
\end{subequations}
Next we discuss the evaluation of $\bar{C}_{nm}$ in Eq.~(\ref{cbar-mat}). 

With the help of Eq.~(\ref{Cnm}), we find 
\begin{align}
\bar{C}^{\alpha}_{nm}(t) 
=\bar{s}^{\alpha}_1(t)A_{nm}^{\dagger}+\bar{s}^{\alpha}_2(t)A_{nm},
\label{barCnm}
\end{align}
where 
\begin{align}
\bar{s}^{\alpha}_{i}(t)= \frac{1}{\hbar}\int_{-\infty}^{\infty} \text{d}\omega 
\,J(\omega) 
n(\omega)\int_0^{\infty}\text{d}\tau\, e^{i\omega \tau}s^{\alpha}_{i}(t-\tau) 
\,.
\label{sbar}
\end{align}
In order to evaluate Eq.~(\ref{sbar}), we choose a sufficiently large 
time interval  $\left[t_i,t_f\right]$ of length $T_{if}=t_f-t_i$ to avoid 
artefacts from the Gibbs effect~\cite{Riley2006} and represent $s^{\alpha}_{i}$ 
in terms of the first $N_k$ terms of its discrete Fourier series, 
\begin{equation}
\label{FourierEq}
s^{\alpha}_{i}(t_i<t<t_f)\approx\sum _{k=-k_{\text{max}}}^{k_{\text{max}}} \mc{F}[s^{\alpha}_{i}](\omega_k)  e^{\imag w_k t}\,.
\end{equation}
Here $w_k=\frac{2 \pi k}{T_{if}}$ are the discrete frequencies with integer 
index $k$, 
$k_{\text{max}}=\frac{N_k-1}{2}$ and the Fourier coefficients $\mc{F}[s^{\alpha}_{i}](\omega_k)$ are defined as 
\begin{equation}
\mc{F}[s^{\alpha}_{i}](\omega_k) = \frac{1}{T_{if}}\int\limits_{t_i}^{t_f} s^{\alpha}_{i}(t) 
e^{-\imag w_k t}\ \text{d}t\,.
\label{G}
\end{equation}        
The expansion in Eq.~(\ref{FourierEq}) allows us to evaluate the integrals  in 
Eq.~(\ref{sbar}). We find 

\begin{align}
\bar{s}^{\alpha}_{i}(t) = & \frac{\pi}{\hbar} \sum _{k=-k_{\text{max}}}^{k_{\text{max}}}J(w_k) 
n(w_k) e^{\imag w_k t}\mc{F}[s^{\alpha}_{i}](\omega_k)\,,
\label{sbarF}
\end{align}
where we employed the identity 
$\int_0^{\infty}\text{d}t e^{\imag\omega 
t}=\pi\delta(\omega)+P(\nicefrac{\imag}{\omega})$ and 
neglected  the principal part. 
We  now return to the master equation~(\ref{MasterEqProj}). 
Substituting Eqs.~(\ref{Cnm}) and~(\ref{barCnm}) in Eq.~(\ref{MasterEqProj}) 
results in  a differential equation for $\rho_{nm}$ 
with time-dependent coefficients $s_i^{\alpha}(t)\bar{s}_j^{\alpha}(t)$.  
In order to be consistent with the Born-Markov approximation we average these 
time-dependent terms 
over the correlation time of the bath $\tau_B$ times a constant factor $f$. In order to make sure that the bath cannot introduce any dynamics  
faster than $\tau_B$, we set $f=10$ in Eq.~(\ref{SecRWA}). This is necessary since the 
Markov approximation 
assumes that the system does not evolve appreciably over this time scale. We 
find 
\begin{align}
\dot{\rho}_{nm}(t)= 
&  \sum_{k,l}\Big\{  
\mathcal{S}_1^{\alpha}(t)\left[A_{nk}\rho_{kl}A_{lm}^{\dagger}-
A_{nk}^{\dagger}A_{kl}\rho_{lm}\right] \notag \\
&\quad +\mathcal{S}_2^{\alpha}(t)\left[A_{nk}^{\dagger}\rho_{kl}A_{lm}-
A_{nk}A_{kl}^{\dagger}\rho_{lm}\right] \notag \\
&\quad +\mathcal{S}_3^{\alpha}(t)\left[A_{nk}\rho_{kl}A_{lm}-
A_{nk}A_{kl}\rho_{lm}\right] \notag \\
& \quad 
+\mathcal{S}_4^{\alpha}(t)\left[A_{nk}^{\dagger}\rho_{kl}A_{lm}^{\dagger}-
A_{nk}^{\dagger}A_{kl}^{\dagger}\rho_{lm}\right] \notag \\
 &  \qquad \Big\}  +\text{H.c.} \,,
\label{eqMaster}
\end{align}
where 
\begin{subequations}\label{SecRWA}
\begin{align}
 \mathcal{S}_1^{\alpha}(t) & = \frac{1}{f\tau_B}\int\limits_{t-f\tau_B }^{t}\text{d}t'
 \bar{s}^{\alpha}_2(t')s^{\alpha}_1(t') \,,\\
 \mathcal{S}_2^{\alpha}(t)& = \frac{1}{f\tau_B}\int\limits_{t-f\tau_B }^{t}\text{d}t'
 \bar{s}^{\alpha}_1(t')s^{\alpha}_2(t')\,, \\
\mathcal{S}_3^{\alpha}(t) & =\frac{1}{f\tau_B}\int\limits_{t-f\tau_B }^{t}\text{d}t' 
\bar{s}^{\alpha}_2(t')s^{\alpha}_2(t')\,,\\
 \mathcal{S}_4^{\alpha}(t) & = \frac{1}{f\tau_B}\int\limits_{t-f\tau_B }^{t}\text{d}t'
 \bar{s}^{\alpha}_1(t')s^{\alpha}_1(t')  \,.
\end{align}
\end{subequations}
By converting Eq.~(\ref{eqMaster}) into an operator-valued 
equation we obtain Eq.~(\ref{eqMaster2}) in the main text.
\section{Closed set of equations of motions for full master equation}\label{FullEqu}
A closed set of differential equations for the components of the the vector 
$\mf{v}=(\langle X^2\rangle ,\langle P^2\rangle,\langle X P+PX\rangle)^\intercal$ with position-momentum 
correlator $D=XP+XP$ can be derived from the full master equation and is given by
\begin{multline}\label{Eqx2full}
\frac{d}{dt}v_1=\omega_0 v_3\\
+\Big\{-i v_1\left[h_1(h^*_2 S_1^{\alpha}+ h_2 S_3^{\alpha})_+ h_1^*(h_2 
S^{\alpha}_2+h^*_2  S^{\alpha}_4)\right]\\
 \frac{-i v_3 +1}{2} \left[h_2(h^*_2 S_1^{\alpha}+h_2 S_3^{\alpha})+ h_2^*(h_2 
S^{\alpha}_2+ h^*_2 S^{\alpha}_4)\right]+\text{h.c.} \Big\}
,
\end{multline}
\begin{multline}\label{Eqp2full}
\frac{d}{dt}v_2=-\frac{\omega^2(t)}{\omega_0}v_3\\
+\Big \{iv_2\left[h_2 (h^*_1  S_1^{\alpha}+h_1 S_3^{\alpha})+ h_2^* (h_1 
S^{\alpha}_2+h^*_1  S^{\alpha}_4)\right]\\
+\frac{iv_3 +1}{2} \left[h_1 (h^*_1  S_1^{\alpha}+h_1  S_3^{\alpha})+h_1^* ( h_1 
S^{\alpha}_2+h^*_1 S^{\alpha}_4)\right]+\text{h.c.} \Big\}
,
\end{multline}
and 
\begin{multline}\label{EqDfull}
\frac{d}{dt}v_3=2 \omega_0 v_2 -2 \frac{\omega^2(t)}{\omega_0}v_1\\
+\Big \{i v_1\left[h_1 (h^*_1  S_1^{\alpha}+h_1 S_3^{\alpha})+ h_1^* (h_1 
S^{\alpha}_2+h^*_1  S^{\alpha}_4)\right]\\
-i v_2\left[h_2 (h^*_2  S_1^{\alpha}+h_2 S_3^{\alpha})+ h_2^* (h_2 
S^{\alpha}_2+h^*_2  S^{\alpha}_4)\right]\\
-\frac{1}{2}\left[(h^{*}_1 h_2+h^{*}_2 h_1)(S^{\alpha}_1+S^{\alpha}_2)+2h_1 h_2 
S^{\alpha}_3  +2 h^{*}_1 h^{*}_2 S^{\alpha}_4\right]\\
-v_3 \left[S^{\alpha}_1-S^{\alpha}_2\right]+\text{h.c.} \Big\}
.
\end{multline}
In case of the HTME, we obtain
\begin{multline}\label{EqHighT}
\frac{d\mf{v}}{dt}= M^{\alpha}_i\mf{v}+A_w \cos{(\omega_L t)}M_t\mf{v}+\mf{I}_{\alpha}\,,
\end{multline}
where 
\begin{align}
&M^{\alpha}_i    \\
& =\left(
\begin{array}{ccc}
 -2 \gamma  \sin^2{(\alpha )} & 0 & \frac{\gamma}{2} \sin{ (2\alpha )} + \omega_0\\
 0 & -2 \gamma  \cos^2{(\alpha )} & \frac{\gamma}{2}  \sin{(2 \alpha )}- \omega_0  \\
\gamma  \sin{(2\alpha )} -2\omega_0 & \gamma  \sin{(2\alpha )} +2\omega_0 & -\gamma  
\end{array}
\right)\notag
\end{align}
determines  the isolated system dynamics,
\begin{equation}
M_t=\left(
\begin{array}{ccc}
 0 & 0 & 0 \\
 0 & 0 & -\omega_0 \\
 -2\omega_0& 0 & 0  \\
\end{array}
\right)
\end{equation}
accounts for the external driving and 
\begin{equation}
\mf{I}_{\alpha}=[2 n(\omega_0)+1] \gamma  \left(
\begin{array}{c}
 \sin^2{(\alpha )} \\
 \cos^2{(\alpha )} \\
 -\sin{(2 \alpha )} \\
\end{array}
\right)
\end{equation}
is the inhomogeneity in Eq.~(\ref{EqHighT}).
\section{Unitary transformation leading to  $\mc{H}^{\alpha}$ \label{SecInterH}}
The unitary transformation relating the total 
Hamiltonian $H^{\alpha}$ to $\mc{H}^{\alpha}$ in Eq.~(\ref{Htrans})  is given by 
\begin{equation}
U=e^{-\frac{i}{\hbar}\frac{\sin{(\alpha)}}{\omega_0}x\sum_r \kappa_r x_r}\,.
\end{equation}
This transformation leaves the position operators of the system and bath remain unchanged, 
i.e., $x_r=U^{\dagger} x_r U$ and $x=U^{\dagger} x U$. However, the 
momentum operators transform as
\begin{subequations}
\begin{align}
U^{\dagger} p_r U=p_r-\frac{\sin{(\alpha)}}{\omega_0}\kappa_r x, \\
U^{\dagger} p U=p-\frac{\sin{(\alpha)}}{\omega_0}\sum_{r=1}^{N}\kappa_r x_r \,.
 \end{align}
\end{subequations}
The term $\mathcal{H}_{\text{shift}}$ in Eq.~(\ref{Htrans}) is given by
\begin{equation}
\mathcal{H}_{\text{shift}}=\left[\frac{1}{2\omega^2_0}\sum_{r=1}^{N} 
\frac{\kappa_r^2}{m_r}x^2-\frac{1}{2 m \omega_0^2}\left(\sum_{r=1}^{N} \kappa_r 
x_r\right)^2\right]\sin^2{(\alpha)}\,.
\end{equation}
\section{Master equation resulting from  $\mc{H}^{\alpha}$ \label{MasterU}}
Following the steps in Appendix~\ref{MasterDerivation}, the transformed Hamiltonian $\mc{H}^{\alpha}$ 
gives rise to the following master equation,
\begin{align}
\dot{\rho}_U= & -\frac{\imag}{\hbar}\left[H^F_{\text{S}}(t),\rhoU\right]\notag \\
& +\Big\{ \mathcal{U}_1^{\alpha}(t) \left[A(t)\rhoU 
A^{\dagger}(t)-A^{\dagger}(t)A(t)\rhoU\right] \notag \\
& \qquad+\mathcal{U}_2^{\alpha}(t) \left[A^{\dagger}(t)\rhoU 
A(t)-A(t)A^{\dagger}(t)\rhoU\right] \notag \\
& \qquad+\mathcal{U}_3^{\alpha}(t) \left[A(t)\rhoU A(t)-A(t)A(t)\rhoU\right] 
\notag \\
& \qquad+\mathcal{U}_4^{\alpha}(t) \left[A^{\dagger}(t)\rhoU 
{A}^{\dagger}(t)-A^{\dagger}(t)A^{\dagger}(t)\rhoU\right] \notag \\
&  \qquad +\text{H.c.}\Big\}\,,
\label{eqMaster3}
\end{align}
where the operators $A$ are the same as in Appendix~\ref{MasterDerivation} and  
\begin{subequations}\label{Ufunc}
\begin{align}
 \mathcal{U}_1^{\alpha}(t) & = \frac{1}{f\tau_B}\int\limits_{t-f\tau_B }^{t}\text{d}t'
 \bar{u}^{\alpha}_2(t')u_1(t') \,,\\
 \mathcal{U}_2^{\alpha}(t)& = \frac{1}{f\tau_B}\int\limits_{t-f\tau_B }^{t}\text{d}t'
 \bar{u}^{\alpha}_1(t')u_2(t')\,, \\
\mathcal{U}_3^{\alpha}(t) & =\frac{1}{f\tau_B}\int\limits_{t-f\tau_B }^{t}\text{d}t' 
\bar{u}^{\alpha}_2(t')u_2(t')\,,\\
 \mathcal{U}_4^{\alpha}(t) & = \frac{1}{f\tau_B}\int\limits_{t-f\tau_B }^{t}\text{d}t'
 \bar{u}^{\alpha}_1(t')u_1(t')  \,.
\end{align}
\end{subequations}
As in Appendix~\ref{MasterDerivation}, $\tau_B$ is the bath correlation time and $f=10$. 
The function $u_{1}$  in Eq.~(\ref{Ufunc}) is given by 
\begin{align}
 u_{1}(t) = -\imag {\sqrt{\frac{\hbar}{2m \omega_0}}} h_2(t) \,, 
\label{s10}
\end{align}
where $h_2(t)$ is defined in Eq.~(\ref{hfunc}) and only depends 
on the parameters of the classical parametric oscillator. 
Furthermore,   $u_2(t)=[u_1(t)]^*$ and 
\begin{align}
\bar{u}^{\alpha}_{i}(t)= \frac{1}{\hbar}\int_{-\infty}^{\infty} \text{d}\omega 
\,J^{\alpha}(\omega) 
n(\omega)\int_0^{\infty}\text{d}\tau\, e^{i\omega \tau}u_{i}(t-\tau) 
\,,
\label{sbarF2}
\end{align}
where $J^{\alpha}$ given  in Eq.~(\ref{J2}). 
By calculating the discrete Fourier transform   $\mc{F}[u_i]$ of $u_i$ and neglecting principle value integrals 
analogous to Eq.~(\ref{sbarF}), Eq.~(\ref{sbarF2}) can be written as 
\begin{align}
\bar{u}^{\alpha}_{i}(t) \approx & \frac{\pi}{\hbar} \sum _{k=-k_{\text{max}}}^{k_{\text{max}}}J^{\alpha}(w_k) 
n(w_k) e^{\imag w_k t}\mc{F}[u_i](\omega_k)\,.
\label{sbarF2f}
\end{align}
Since  $u_{1}\propto h_2$ and $u_{2}\propto h_2^*$, Eq.~(\ref{sbarF2f}) shows that the system-bath interaction 
is determined by the spectral density $J^{\alpha}(w_k)$ 
evaluated at the characteristic frequencies $\omega_k=\omega_c$ of the classical harmonic oscillator solutions.

%\bibliography{library,kiffner,books}
%merlin.mbs apsrev4-1.bst 2010-07-25 4.21a (PWD, AO, DPC) hacked
%Control: key (0)
%Control: author (8) initials jnrlst
%Control: editor formatted (1) identically to author
%Control: production of article title (-1) disabled
%Control: page (0) single
%Control: year (1) truncated
%Control: production of eprint (-1) disabled
%

\end{document}